\documentclass[10pt]{article}
\usepackage{graphicx,amssymb,wasysym}

\begin{document}

\title{Empirical analysis of the Lieb-Oxford bound in ions and molecules}
\author{Mariana M. Odashima and K. Capelle\\
Instituto de F\'{\i}sica de S\~ao Carlos\\
Universidade de S\~ao Paulo\\
Caixa Postal 369, S\~ao Carlos, 13560-970 SP, Brazil}

\date{\today}

\maketitle

\begin{abstract}
Universal properties of the Coulomb interaction energy apply to all 
many-electron systems. Bounds on the exchange-correlation energy, in 
particular, are important for the construction of improved density functionals.
Here we investigate one such universal property --- the Lieb-Oxford lower 
bound --- for ionic and molecular systems. In recent work [J. Chem. Phys. 
{\bf 127}, 054106 (2007)], we observed that for atoms and electron liquids 
this bound may be substantially tightened. Calculations for a few ions and 
molecules suggested the same tendency, but were not conclusive due to the
small number of systems considered.  Here we extend that analysis to many
different families of ions and molecules, and find that for these, too, the
bound can be empirically tightened by a similar margin as for atoms and
electron liquids. Tightening the Lieb-Oxford bound will have consequences for the
performance of various approximate exchange-correlation functionals.
\end{abstract}


\newcommand{\be}{\begin{equation}}
\newcommand{\ee}{\end{equation}}
\newcommand{\bi}{\bibitem}
\newcommand{\la}{\langle}
\newcommand{\ra}{\rangle}
\newcommand{\ua}{\uparrow}
\newcommand{\da}{\downarrow}
\newcommand{\bea}{\begin{eqnarray}}
\newcommand{\eea}{\end{eqnarray}}
\renewcommand{\r}{({\bf r})}
\newcommand{\rp}{({\bf r'})}

\section{\label{intro}Introduction}

The Lieb-Oxford lower bound \cite{lopaper} on the exchange-correlation 
($xc$) energy is an exact quantum-mechanical property of any 
Coulomb-interacting system. Originally, this bound was obtained in an 
investigation of a very fundamental problem of stability of matter \cite{lieb}. 
An important practical application of the bound is as a constraint in the
construction of approximate exchange-correlation functionals for
density-functional theory (DFT).

In DFT \cite{kohnrmp,dftbook,parryang}, many-body effects on electronic 
structure are included by means of an approximate $xc$ functional. 
The Lieb-Oxford bound is one of the key ingredients in the construction 
of nonempirical $xc$ functionals \cite{perdewreview}, where it is usually 
written as
\be
E_{xc}[n] \ge -C \int d^3r\,n^{4/3}, 
\label{eq:ineq1}
\ee
where $C$ is a positive universal constant whose exact value is unknown. 
A first estimate for the maximum value of this constant, $C_{L}=8.52$ 
\cite{lieb} was later substituted \cite{lopaper} by the tighter value 
$C_{LO}=1.68$, which is adopted in most current work. 

With this value of $C$, the bound is satisfied for  
instance by the local-density approximation (LDA) \cite{pw92,vwn}, the 
generalized-gradient approximations (GGAs) PW91 \cite{pw91} and PBE 
\cite{pbe}, and the TPSS meta-GGA \cite{tpss}.  On the other hand, earlier 
GGAs \cite{pw86} and meta-GGAs \cite{pkzb} and other functionals containing 
fitting parameters \cite{b88,lyp,revPBE,becke07} are not guaranteed 
to satisfy the bound for all possible densities. The success of some these
latter functionals shows that, while the bound is doubtlessly obeyed by the
exact functional, satisfaction is not a necessary condition for good 
performance in practice. 

In terms of the local-density approximation to the exchange energy,
\be
E_x^{LDA}[n]= -\frac{3}{4}\left(\frac{3}{\pi}\right)^{1/3} \int
d^3r\,n^{4/3} \;,
\ee
Eq.~(\ref{eq:ineq1}) becomes
\be
E_{xc}[n] \ge \lambda E^{LDA}_x[n],
\label{eq:ineq2}
\ee 
where $\lambda=\frac{4}{3}\left(\frac{\pi}{3}\right)^{1/3}\,C$. The value
corresponding to the Lieb-Oxford estimate $C_{LO}=1.68$ is 
$\lambda_{LO}=2.275$. The exact universal $\lambda$ and $C$ are
unknown, but their existence is guaranteed by the Lieb-Oxford bound,
and their values are constrained by $\lambda \leq \lambda_{LO}$ and
$C \leq C_{LO}$, respectively.

More recently, Chan and Handy numerically obtained an additional small 
reduction of this upper limit of $C$ to $C_{CH}=1.6358$, implying
$\lambda_{CH}=2.215$ \cite{handylo}.
The present authors \cite{looc} empirically investigated the bound for
atoms, model Hamiltonians and the electron liquid, as well as for a very
restricted number of ions and molecules. It turned out that for all systems 
investigated there even the tightest previous formulation of the bound, 
employing $C_{CH}=1.6358$, is still substantially too generous. Away from
unphysical limiting cases of model Hamiltonians, $C_{conjec}=1.0$ was
found to be appropriate for all stable real systems investigated there. 
In view of the impact a reduction of $C$ (or, equivalently, of $\lambda$) 
may have on current density functionals, this observation requires further 
investigation.

In the present paper, we enlarge the data base of the empirical investigation 
to include diverse families of ions and molecules, in order to 
accumulate further evidence and to be able to quantify the possible 
degree of tightening for molecular systems. To this end, we
use high-precision Quantum Monte Carlo (QMC), configuration interaction
(CI) and exact exchange (EXX) data to evaluate the ratio
\be
\frac{E_{xc}[n]}{E_x^{LDA}[n]}=:\lambda[n],
\label{eq:lambda}
\ee
for different families of ions and molecules, including first and
second-row dimers, and various hydrides and oxides.

This paper is organized as follows: in Sec.~\ref{ions} we discuss various
ions from the $Be$, $Li$, $C$ and $O$ isoelectronic series, as well as 
monovalent cations, comparing their behaviour with data for the $He$
isoelectronic series and atoms obtained in Ref.~\cite{looc}. 
Section \ref{molecules} deals with different classes of molecules,
and Sec.~\ref{analysis} provides a comparative analysis of the emerging 
trends. Section~\ref{concl} contains our conclusions.

\section{Ions}
\label{ions}

\begin{figure}[t]
\includegraphics[width=8cm,height=6cm]{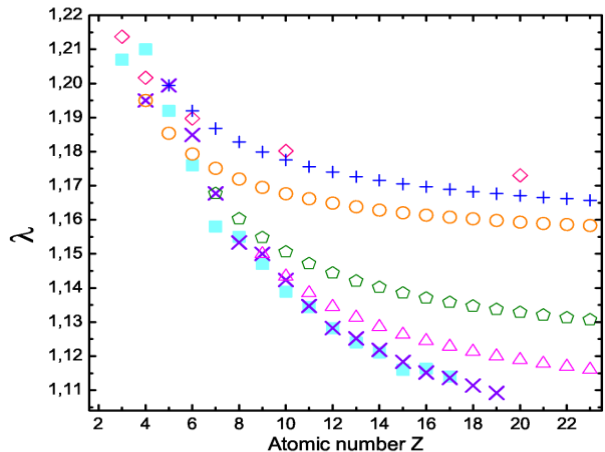}
\centering
\caption {\label{fig1} Value of the ratio $\lambda$, defined in
Eq.~(\ref{eq:lambda}), as a function of atomic number $Z$ for 
different ions. Ions isoelectronic to He (diamonds), Be (plus), Li 
(circles), C (pentagons) and O (triangles), and monovalent cations 
(crosses) are shown. Values for neutral atoms (squares)
\cite{looc} are included for comparison.}
\end{figure}

Results from evaluating $\lambda[n]$ for ions from the $He$, $Be$, $Li$, $C$ 
and $O$ isoelectronic series, and for monovalent cations are displayed in
Fig.~\ref{fig1}.

According to Eq.(\ref{eq:lambda}), in order to calculate the parameter 
$\lambda[n]$ we need the exact exchange-correlation energy $E_{xc}[n]$ and 
the LDA exchange energy $E_x^{LDA}[n]$. 
Ionic correlation energies were taken from Ref.~\cite{froese}, except
for the $He$ series, for which we employed the DMC $xc$ energy given by
Huang and Umrigar \cite{huangumrigar}.
To these near-exact correlation energies we added EXX energies calculated 
with the OEP method \cite{oep,opmks}
to simulate the exact $xc$ energy. Small differences between exact exchange 
in DFT and in Hartree-Fock theory are negligible on the scale of effects
we are after. The values for $E_x^{LDA}$ were obtained \cite{opmks} 
from the LDA exchange functional and evaluated at self-consistent 
LDA(VWN)\cite{vwn} densities. 

Figure~\ref{fig1} shows that ionic $\lambda$ values are similar to those of atoms,
and far away from the theoretical upper limits. $\lambda[n]$ decreases as the 
ionic charge increases and the density distribution becomes more compact.
Chemically similar ions (same number of electrons) fall on common $\lambda(Z)$
curves, displaying systematic trends as a function of atomic number and
total charge.
Monovalent cations have charge distributions that are very similar to
neutral atoms, in particular for larger $Z$, and consequently have almost
the same $\lambda$ values.

\section{Molecules}
\label{molecules}

\begin{figure}[t]
\centering
\includegraphics[width=8cm,height=6cm]{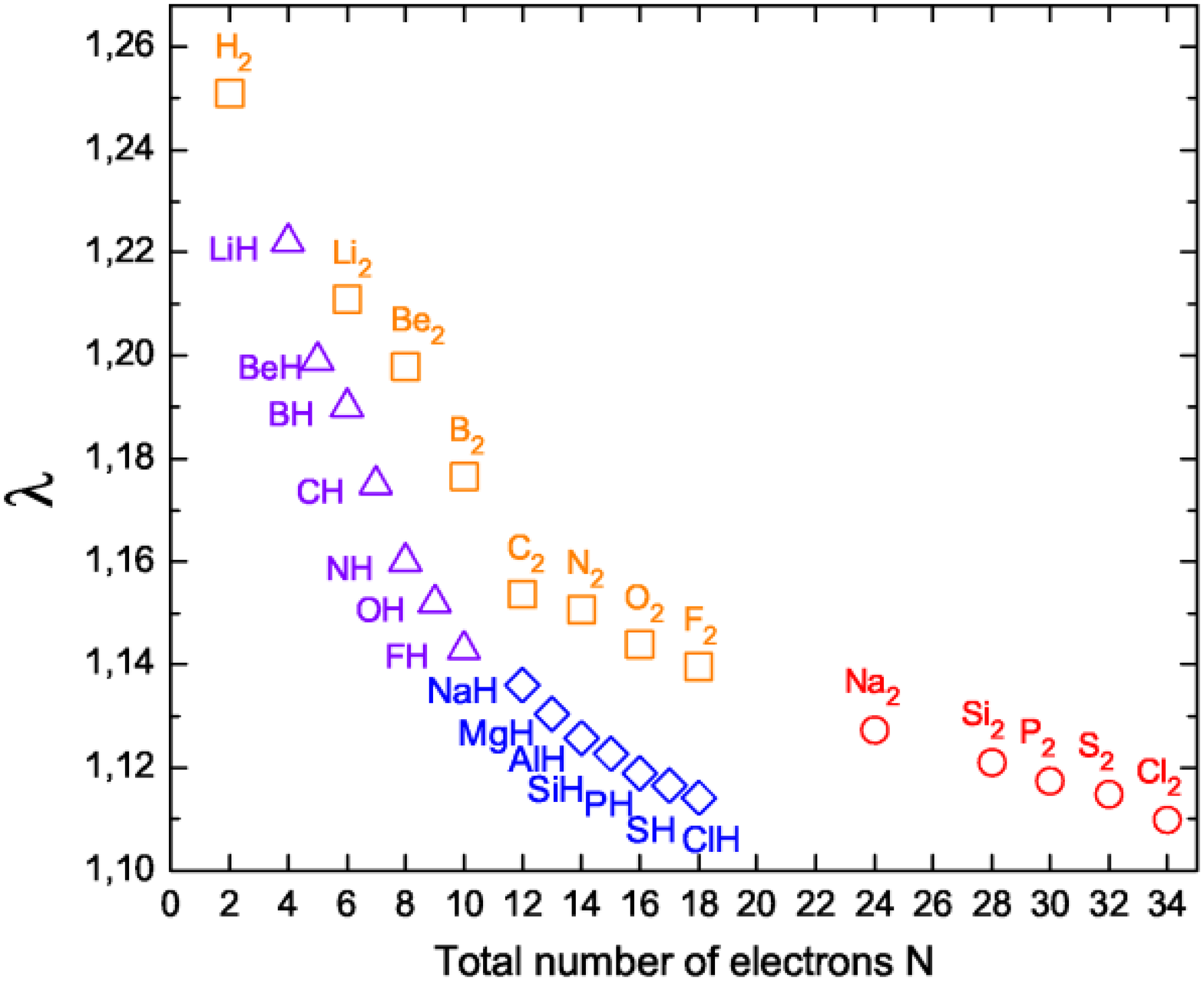}
\caption {\label{fig2} Energy ratio $\lambda$ as a function of the 
total number of electrons $N$, for first-row dimers (squares), 
second-row dimers (circles), first-row hydrides (triangles)
and second-row hydrides (diamonds).}
\end{figure}

\begin{figure}[t]
\centering
\includegraphics[width=8cm,height=6cm]{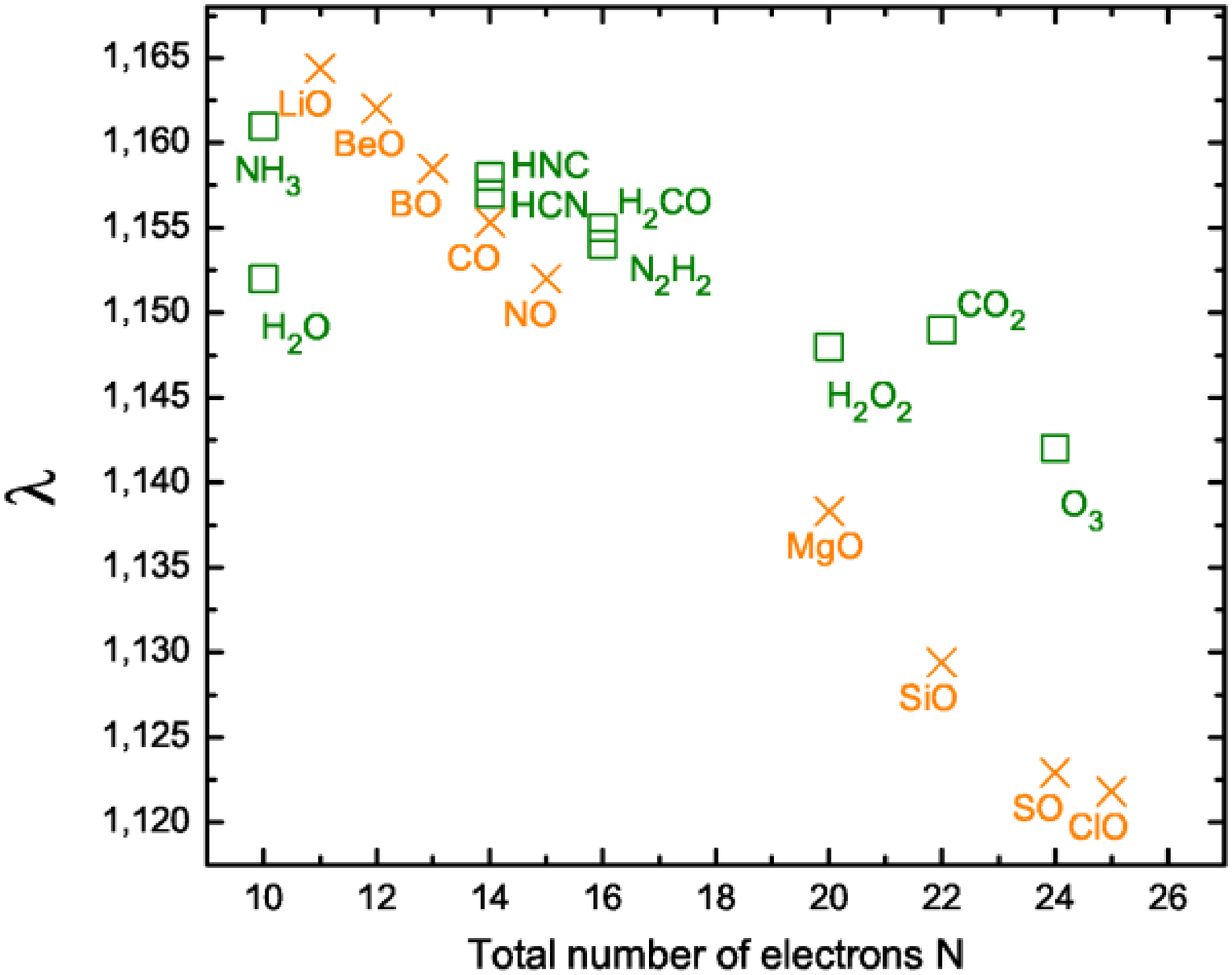}
\caption {\label{fig3} Energy ratio $\lambda$ as a function of the
total number of electrons $N$, for first and second-row oxides (crosses)
and a set of miscellaneous other molecules for which near-exact $E_{xc}$
data are available.}
\end{figure}

Figure~\ref{fig2} reports $\lambda$ values for a set of dimers and hydrides. 
Figure~\ref{fig3} extends this comparison to oxides and a variety of other 
molecules for which near-exact correlation energies are available.

For the first-row dimers we employed highly-accurate diffusion Monte Carlo 
(DMC) correlation energies from Ref.~\cite{mantenluchow} ($H_2$) and
Ref.~\cite{filippiumrigar} ($Li_2$ to $F_2$). Correlation energies for the
second-row dimers are from Ref.~\cite{oneillgill}, where they were extracted
from experimental dissociation energies, corrected for zero-point 
oscillations, by subtracting Hartree-Fock total energies.
For the first-row hydrides we used DMC correlation energies extracted 
from Ref.~\cite{luchowanderson}, while correlation energies of second-row 
hydrides were obtained from experimental dissociation energies in 
Ref.~\cite{vosko}. For first and second-row oxides, as well as for the
other molecules shown in Fig.~\ref{fig3}, reference correlation 
energies are taken from \cite{oneillgill} and \cite{vosko}. 

These correlation energies were combined with LDA exchange energies and
Hartree-Fock exchange (again, small differences between exact exchange 
in DFT and in Hartree-Fock theory are negligible on the scale of effects we 
are after) to calculate the ratio 
$\lambda[n]=(E_x^{\rm exact}+E_c^{\rm exact})/E_x^{LDA}$. 
The Hartree-Fock and LDA calculations were performed with the Gaussian 03 
package \cite{gaussian}. These calculations used the 6-311G+(3df,2p) basis sets for the
first-row dimers and hydrides, and the cc-pVTZ basis sets for the second-row
dimers and hydrides, and all oxides and remaining molecules. Geometries for 
the Hartree-Fock and LDA calculations were taken from the references providing 
the correlation energies. For some molecules the geometries were 
not supplied together with the correlation energies \cite{mantenluchow}. In these cases we 
performed our own geometry optimization on the MP2/cc-pVTZ level, followed
by single-point calculations using the basis sets and methods indicated above.

Note that in Figs.~\ref{fig2} and \ref{fig3} we display results 
as a function of the number of electrons $N$ of each molecule, instead of 
the atomic number $Z$ used for atoms and ions. Nevertheless, the tendencies
observed for ionic atoms clearly persist for neutral molecules: $\lambda(N)$ 
decreases for larger (heavier) systems and increases for more rarefied 
electronic distributions; chemically similar families of
systems fall on common $\lambda(N)$ curves; and values of $\lambda(N)$ are 
far from $\lambda_{LO}$. 

The tendency of $\lambda$ to increase for more rarefied electronic 
distributions (lower average electron density) was already observed in 
the electron gas and for Hooke's atom \cite{looc}. To further investigate 
this tendency for molecules, we also considered
the variation of $\lambda$ as a function of the distance between 
the nuclei of the $H_2$ molecule. Approximate correlation energies of
$H_2$ as a function of bond length are available from Ref.~\cite{baerends}.
From these data we find that for $H_2$ at the equilibrium bond length 
$\lambda(R_e=1.401a.u.)=1.230$ (only slightly different from the value $1.251$,
obtained for the same system from DMC), whereas for stretched $H_2$ this 
changes to $\lambda(R_e=5.0a.u.)=1.409$. Even this enhanced value of $\lambda$, 
though, is still way below the maximum $\lambda_{LO}=2.275$.

\section{Analysis}
\label{analysis}

The $\lambda[n]$ functional, introduced in Ref.~\cite{looc}, has two 
key properties. One is that by definition it measures the inverse weight 
of LDA exchange relative to exact exchange and correlation: The 
larger $\lambda[n]$, the smaller is $|E_x^{LDA}[n]|$ relative to 
$|E_{xc}^{\rm exact}[n]|$. $\lambda[n]$ therefore serves to characterize
the importance of nonlocal exchange and of correlation in a given system
or class of systems, indicating when $E_x^{LDA}[n]$ may be a good starting
point and when more refined approaches are required.

The other key property is that the maximum of $\lambda[n]$ for a given
class of systems provides a class-specific Lieb-Oxford bound. If these classes
are large enough, such bounds may prove very useful in practice. Moreover,
the maximum of $\lambda[n]$ across all investigated systems provides an
empirical approximation to the universal constant $\lambda \leq
\lambda_{LO}$ whose existence is guaranteed by the Lieb-Oxford bound 
(\ref{eq:ineq2}), but whose exact value remains unknown.

From the point of view of these two key properties, three common aspects of 
all $\lambda$ data presented in this paper deserve special attention:

(i) $\lambda$ decreases as the density distribution becomes more compact
({\em e.g.}, for increasing $Z$ in isoelectronic series of ions).
Larger $\lambda[n]$ arise for systems with spread-out rarefied density
distribution. This trend is consistent with the characterization of very
low-density systems as strongly correlated, which is rigorously correct
for uniform electron liquids.
 
(ii) Chemically similar systems are characterized by common trends of
$\lambda$, suggesting that $\lambda$ may be a useful parameter to 
characterize and classify systems, and that $\lambda[n]$ may be a useful
ingredient in the construction of refined $xc$ functionals.

(iii) All values of $\lambda$ obtained are much smaller than the maximum 
value $\lambda_{LO}$, averaging at around half of $\lambda_{LO}=2.275$. 
The new ionic and molecular data reported here are fully consistent with
the conclusion obtained in Ref.~\cite{looc}, and reinforce the conjecture
that the Lieb-Oxford bound can be substantially tightened. The empirical 
value suggested in Ref.~\cite{looc} for real systems ({\em i.e.} excluding 
unrealistic extreme limits of model Hamiltonians), $\lambda_{conjec}\approx 
1.35$ (corresponding to $C_{conjec}\approx 1.0$) still covers all new systems
investigated here, for which $\lambda$ never exceeds $1.26$.

If we include unstable systems, such as $H_2$ stretched to nearly four times 
its equilibrium distance, $\lambda[n]$ exceeds the value conjectured for
stable systems at equilibrium ($1.409$ versus $1.35$), which agrees with 
the trend observed for Hooke's atom in Ref.~\cite{looc}. Even at equilibrium,
$H_2$ has with $\lambda=1.251$ the largest value found for any system 
investigated here, but this value is still clearly below the conjectured
limit $\lambda_{conjec}\approx 1.35$.

\section{Conclusions}
\label{concl}

Numerical results for the Lieb-Oxford parameter $\lambda[n]$, based on
high-quality input data for the correlation energy $E_c[n]$, confirm the
expectation of Ref.~\cite{looc} that for all realistic systems the upper 
limit $\lambda_{LO}=2.275$ is too generous, by almost a factor of two. 
This is the same conclusion obtained in Ref.~\cite{looc}, but considerably
strengthened by the present analysis covering many different types of
ions and molecules not considered there. 

Since we have not found a single exception to the emerging trend 
\cite{footnote}, our analysis suggests to pursue the following further 
investigations:

(i) Mathematical scrutiny of the Lieb-Oxford bound, which may not be
the strongest possible bound for arbitrary Coulomb systems. Moreover, even
if application to arbitrary systems ({\em i.e.} a truly universal bound) 
should require placing $\lambda$ at or near $\lambda_{LO}=2.275$, 
"realistic" systems, away from unphysical limits of system parameters,
may allow a much smaller value of the upper limit. It remains to be
clarified how to characterize such realistic systems or parameter values
in a way independent of $\lambda$ itself.

(ii) Exploration of consequences a reduced value of $\lambda$ has for
the performance of common approximate density functionals. This exploration
is aided by an investigation of the emerging systematic behaviour of 
$\lambda[n]$ accross different families of molecules, to obtain analytical 
or numerical models of the functional $\lambda[n]$.

Work on these issues is in progress in our group.\\

{\bf Acknowledgments}
This work was supported by FAPESP and CNPq.


\end{document}